\pdfoutput=1

\documentclass[twoside,letterpaper]{soups} 
\usepackage{balance}  
\usepackage{booktabs}

\usepackage{graphicx}

\usepackage{tabularx}
\usepackage{float}
\usepackage{multicol}
\usepackage[flushleft]{threeparttable}

\usepackage[table]{xcolor}
\usepackage{enumitem}
\usepackage{caption}
\usepackage{framed}

\usepackage[hyphens]{url}      

\emergencystretch=\maxdimen
\renewcommand\footnotemark{}

\begin{document}
%
\conferenceinfo{Symposium on Usable Privacy and Security
  (SOUPS)}{2016, June 22--24, 2016, Denver, Colorado.}
\CopyrightYear{2016} 

\title{User Attitudes Toward the Inspection of Encrypted Traffic\thanks{\scriptsize \hspace{-11pt} This work was supported by Sandia National Laboratories, a 2014 Google Faculty Research Award, and the National Science Foundation under Grant No. CNS-1528022.
Any opinions, findings, and conclusions or recommendations expressed in this material are those of the author(s) and do not necessarily reflect the views of the National Science Foundation.
Sandia National Laboratories is a multi-program laboratory managed and operated by Sandia Corporation, a wholly owned subsidiary of Lockheed Martin Corporation, for the U.S. Department of Energy's National Nuclear Security Administration under contract DE-AC04-94AL85000.}}

\numberofauthors{1}
\author{
Scott Ruoti\raisebox{7pt}{$\ast$}\raisebox{7pt}{$\dagger$}, Mark O'Neill\raisebox{7pt}{$\ast$}\raisebox{7pt}{$\dagger$}, Daniel Zappala\raisebox{7pt}{$\ast$}, Kent Seamons\raisebox{7pt}{$\ast$}\\
       \affaddr{Brigham Young University\raisebox{4pt}{$\ast$}, Sandia National Laboratories\raisebox{4pt}{$\dagger$}}\\
       \email{ruoti@isrl.byu.edu, mto@byu.edu, \{zappala, seamons\}@cs.byu.edu}
}

\maketitle

\begin{abstract}
This paper reports the results of a survey of 1,976 individuals regarding their opinions on TLS inspection, a controversial technique that can be used for both benevolent and malicious purposes.
Responses indicate that participants hold nuanced opinions on security and privacy trade-offs, with most recognizing legitimate uses for the practice, but also concerned about threats from hackers or government surveillance.
There is strong support for notification and consent when a system is intercepting their encrypted traffic, although this support varies depending on the situation.
A significant concern about malicious uses of TLS inspection is identity theft, and many would react negatively and some would change their behavior if they discovered inspection occurring without their knowledge.
We also find that a small but significant number of participants are jaded by the current state of affairs and have lost any expectation of privacy.
\end{abstract}

\section{Introduction}
In early 2013, one of the authors received an email from a former student who expressed serious concerns after becoming aware that his employer was inspecting its employees' encrypted Internet traffic in order to protect the network from attackers.
Though he was himself employed in the computer security industry, he expressed surprise and anger that this could happen, and also mentioned his serious concerns about the potential for employees to disclose personal information without being aware that their data was visible to their employer. 
He questioned whether this practice was legal and whether it was ethical to do this without notifying employees in advance.

In fact, it is common practice for companies to inspect employees' encrypted traffic to filter malware and viruses, prevent the leak of intellectual property, and block harmful websites~\cite{bluecoatproxysg,symantecwebgateway,paloalto}.
This inspection is usually accomplished with a network device that acts as a TLS/SSL proxy, sitting in the middle of the communication between a browser and web server where it can intercept, decrypt, inspect, then re-encrypt and forward on the user's traffic to its original destination.
This is all accomplished without any visible notification to the user that their encrypted traffic is being inspected.

While security experts overwhelmingly view the inspection of encrypted traffic by attackers and governments as undesirable, the practice of businesses and organizations inspecting their \textit{own} encrypted traffic in order to secure their \textit{own} network and intellectual property is more controversial.
Many experts are alarmed by any use of a TLS proxy because it is deceptive; users' browsers continue to inform them they have a secure connection to the server, even though this is not the case.
Most research in the literature treats all TLS proxies as undesirable and actively tries to prevent their use~\cite{clark2013sok}.
Still, a smaller number of researchers are investigating how the malicious uses can be prevented while still allowing for benevolent use of encrypted traffic inspection by businesses and organizations~\cite{mcgrew2012internetdraft,loreto2014internetdraft}.

While the opinions of businesses and security experts regarding the inspection of encrypted traffic are known, no prior work has measured general (i.e., non-expert) user attitudes and preferences toward the inspection of encrypted traffic.
To better understand users' perspectives on this issue, we surveyed 1,976 people across two surveys regarding their opinions of TLS proxies and their use in inspecting encrypted traffic.\footnote{The full data from both surveys is available at \url{https://soups2016.isrl.byu.edu/}.}
The results of the first survey of 1,049 individuals showed a surprising
willingness by participants to accept the inspection of encrypted traffic, provided they are first notified.
Based on the results of the first survey, we conducted a second survey of 927 individuals to further explore user attitudes towards inspection of encrypted traffic in specific situations.

Our contributions from these surveys include the following insights:

\begin{itemize}

\item
User opinions toward TLS proxies and the inspection of encrypted traffic are nuanced.
Many express concerns about privacy and identity theft from hackers (75.8\%) or government surveillance (70.9\%).
Yet there is broad, general acceptance of TLS proxies when used by employers, schools, etc (71.7\%).

\item
Most participants indicated support for the inspection of encrypted traffic as long as they were first notified of it (90.7\%).
Likewise, participants indicated strong support for legislation requiring notification or consent (83.2\%).

\item
When asked about specific situations in which TLS proxies might be used (e.g., at work, at school, at a caf\'e, or at home), support for TLS proxies ranges from 65\% to 90\% of participants (including those who want notification or consent).  
Support for inspection of encrypted traffic without notification or consent is strongest at elementary schools (45.9\%) and at businesses when employees are using company-provided computers (47.9\%).
Participants generally favor consent in cases when they feel in control (at home, free WiFi, their own device at work) versus
notification when an organization is in control (public library, school, company computer).
In nearly all the scenarios we posed, only a small minority of the participants indicated that using TLS proxies is not acceptable.
The one exception is government surveillance, in which case 47.5\% say that this is not acceptable.

\item
Many users would have a negative opinion if they discovered that the owner of their network used a TLS proxy without prior notification and/or consent (60.8\%), though for some (34.2\%) it would depend on who the owner was and how they were using the technology.
Some would change their behavior on the network, either discontinuing to use it (17.2\%) or changing which sites they visited (6\%).

\item
We identify personas based on participants' responses regarding TLS proxies: pragmatic (76.5\%), privacy fundamentalist (17.0\%), jaded (5.0\%), and unconcerned (1.0\%).
Jaded participants are interesting in that their opinions regarding privacy and security align with the privacy fundamentalist persona, but their practices align with the unconcerned persona.
This dichotomy stems from the fact that these users feel that regardless of what steps they take, they are powerless to prevent compromise of their online information, and so choose to not do anything to protect themselves.

\end{itemize}

While several of our findings might seem intuitive, it is important to ground intuitions in data,
and this paper provides the first survey of user opinions on this topic.
In addition, participants showed a high level of engagement in the survey, notwithstanding
the complexity of the topic. Many users shared in-depth analysis of trade-offs in open responses, demonstrating that they
care deeply about this issue.
User attitudes toward TLS proxies provide an important data point along the spectrum of discussion that is currently taking place regarding who should have access to encrypted information.


\section{Background}
The focus of our surveys is on user attitudes towards the inspection of encrypted traffic (i.e., HTTPS),
specifically with the use of TLS proxies.
In this section we provide technical details regarding TLS proxies.
We also discuss real-world examples of how TLS proxies are used.
Finally, we present related work on measuring user sentiment towards online privacy.

\subsection{TLS Proxies}
When a web browser attempts to validate the identity of a website, it relies on certificate authorities (CAs) that digitally sign certificates vouching for the identity of servers.
Web browsers authenticate a site by validating a chain of trust from the site's certificate back to one of a set of trusted root certificates. 
These certificates comprise the {\em root store} and are typically bundled with the operating system or browser.

This validation system is currently being co-opted by the use of TLS proxies that act as a man-in-the-middle (MitM) for TLS connections.
A TLS proxy can issue a {\em substitute certificate} for any site the user visits, so that the user establishes an encrypted connection to the proxy rather than the desired web site.
The proxy can then decrypt and monitor or modify all user traffic, before passing it along via a second encrypted channel to the desired web site.
For example, when a user attempts to create a secure connection to Amazon by requesting Amazon's certificate, the proxy intercepts this request, generates a certificate for Amazon, and sends this substitute certificate back to the user's machine.
The user's machine will then create a secure connection to the proxy (instead of Amazon) and send all of its data to the proxy, which has full access to it before forwarding it on to Amazon's servers.

TLS proxies can be used for both benevolent and malicious purposes. 
Some companies use TLS proxies to filter malware and viruses, prevent the leak of company secrets and intellectual property, block harmful websites, or catch malicious insiders.
However, less scrupulous companies, government agencies, crime organizations, and others may also use proxies to steal a user's sensitive data, conduct surveillance, or commit identity theft.
Currently, browsers and users have no method for distinguishing between benevolent and malicious TLS proxies, and the user is entirely unaware that an organization or attacker is intercepting encrypted traffic.
Even when a TLS proxy is present, the browsers displays a reassuring lock icon that could mislead users to assume they are communicating securely with the website.

To avoid browser warnings that self-signed substitute certificates would trigger, TLS proxies generate substitute certificates signed by a CA that the user's machine trusts. This can be done in several ways:

\begin{itemize}
	\item Purchasing an intermediate certificate authority certificate.
	\item Installing a new trusted root certificate on the user's machine. This can be done either by businesses (e.g., custom system image, manual installation, enterprise PKI system) or by malware.
	\item Including the certificate on a device's root store when it is manufactured. Nokia was recently found to be using TLS proxies on mobile devices \cite{nokia} and Lenovo has pre-installed software using a TLS proxy on its laptops \cite{lenovo}.
	\item Controlling a root certificate authority. Some governments are in this position, and evidence suggests that even when governments do not own the root they can coerce authorities into granting them certificates for domains they do not own \cite{marlinspike2011ssl,coercion}.
	\item Stealing existing root and intermediate certificate authority certificates \cite{marlinspike2011ssl,postmortem}.
\end{itemize}

\subsection{Real-world Examples}
There are a variety of real-world scenarios, ranging from suspicious to malicious, where inspection of encrypted traffic is documented as having occurred. 

Reports have notified the public that both Nokia and Lenovo used TLS proxies to decrypt customer (not employee) traffic for reasons other than security. 
Nokia decrypted cell phone data, allegedly to improve performance on their cellular network~\cite{nokia}. 
Some Lenovo laptops came with third party software that inserted ads into encrypted data~\cite{lenovo}. 
Weaknesses in the adware implementation left users vulnerable to attack from malicious outsiders.
Public outcry caused both companies to stop accessing encrypted traffic.

Government surveillance has been reported to use similar methods
\cite{soghoian2012certified}. 
A report from 2011 showed that Iran monitored 300,000 citizens online using a stolen certificate from Diginotar, a company that is trusted to certify legitimate websites~\cite{postmortem}.

Two recent measurement studies show that TLS proxies account for about 1 in 250 encrypted connections on the web \cite{huang2014analyzing,o2014tls}.  
The vast majority of these monitored connections are for benevolent purposes, but a small percentage appear to be adware, grayware, and otherwise suspicious activity.

The TLS proxy capability is essentially a backdoor into the current web authentication system. This backdoor has benevolent uses to strengthen the security of users and organizations, and a majority of users support their use. As with any backdoor, it's very existence increases the attack surface that can be exploited by attackers. For example, a recent study of client-side TLS proxies used in personal firewalls and parental filters discovered implementation flaws in a number of products that open the user to attack and weaken their security~\cite{de2016killed}.

\subsection{Related Work}
There have been prior studies that survey user's attitudes about their online security and privacy.
Still, no prior study has looked specifically at user attitudes toward the inspection of encrypted traffic.

McDonald and Cranor~\cite{mcdonald2010americans} used interviews and a survey to explore user's knowledge and perception 
of online behavioral advertising practices. 
They discuss the potential chilling effect of these practices based on 40\% of the users that self-reported they would change their behavior if they learned advertisers were collecting data. 
Similarly, users reported in our survey that they would change their behavior if they learned that their encrypted data was being inspected.

Ur et al.~\cite{ur2012smart} also studied user opinions about online behavioral advertising by conducting 
48 semi-structured interviews with non-technical users. Similar to our work, they found users had nuanced opinions about the 
trade-offs for a technology that was both useful and privacy invasive. 
They determined that users were not receiving effective notice and choice mechanisms. 
Our surveys reveal a strong desire for notification and choice regarding the inspection of encrypted traffic.

Shay et al.~\cite{shay2014my} surveyed users via Amazon Mechanical Turk about their attitudes and experiences with compromised email or social networking sites.
They found that many respondents gave high quality responses to open response questions and 
discussed implications for security mechanism designers. 
Likewise, our work has significance for the designers of mechanisms to inspect encrypted traffic.

Anton et al.~\cite{anton2010internet} surveyed users in 2008 to see if their attitudes on privacy concerns had changed from the same survey administered in 2002. They found that the top three concerns of U.S. users were information transfer, notice/awareness, and information storage. While the top three concerns had not changed, their level of concern had risen.
The top three concerns for European users were the same but in a different order; notice/awareness came in third place.
Concerns for notice/awareness are important to both groups, and was a prominent factor in our surveys.


Woodruff et al.~\cite{woodruff2014would} examined how well users' classification by the Westin Privacy Segmentation Index predicted their actual behavior.
They found that although many participants were classified as privacy fundamentalists, their actions in hypothetical situations were not consistent with this classification.
Similarly, while we group participants into personas with names similar to the Westin categories, we do so by looking at how participants indicate they would react to hypothetical situations and not using any of Westin's several privacy indexes.

\section{First Survey -- Methodology}
In February 2014, we conducted the first online survey using the Amazon Mechanical Turk (MTurk) crowdsourcing service. 
We gathered responses on Wednesday, February 12, 2014 between 7:50 AM and 5:22 PM (PST). Each participant could take the survey once and received \$1 USD as compensation upon completing the survey.
In total 1,262 people completed the online survey.
The survey was approved by our Institutional Review Board and is contained 
in Appendix~\ref{appx:study-one}.

\subsection{Instructing Participants}

Before conducting this survey, we felt it was unlikely that most people would be aware of TLS proxies (an assumption that was upheld by our results).
This presented a dilemma: either we would need to only survey individuals who were already aware of TLS proxies or we would need to instruct participants about TLS proxies.
Both of these options have significant drawbacks.
Limiting the survey to individuals with pre-existing knowledge regarding TLS proxies would likely limit us to participants with highly technical backgrounds, thus failing to gather information about broader opinions related to the inspection of encrypted traffic.
On the other hand, instructing participants on TLS proxies has the risk of unintentionally biasing them one way or another, and requires them to answer questions about a subject they potentially just learned about.

Because our research goal was to survey broad opinions regarding the inspection of encrypted traffic, we preferred not to limit our population to the small fraction of users who are already aware of this issue.
Instead, we chose to accept the limitations related to instructing participants about TLS proxies and survey as many participants as possible.
For our goals, this was preferable to ignoring the opinions of a large portion of users.

To address the risks related to instructing participants on an issue and then surveying them, we spent considerable effort and time crafting our description of TLS proxies.
Our goals were to (1) give a simple and concise overview of how TLS proxies are used to inspect encrypted traffic, and (2)
present participants with a fair and unbiased description of how the inspection of encrypted traffic could be used for both benevolent and malicious purposes.

In preparation for writing the description of TLS proxies, we examined the literature and observed that existing descriptions of TLS proxies were not neutral in tone and would unduly bias participants.
We talked with businesses that sell proxies (i.e., Blue Coat, Symantec) and read opinions from privacy advocates to better understand both sides' opinions.
Based on the information in these sources, we composed a draft of our description of TLS proxies, focusing on using language that was informative and neutral in tone, allowing participants to form their own opinions.
Our team of researchers, which included members who are fundamentally opposed to TLS proxies and members who accept their benevolent uses, iterated on this description until all members were satisfied with its wording.

We then tested this description using a convenience sample of six individuals from our university who were not a part of our research group to ensure it was balanced and understandable.
Based on feedback from the convenience sample, we made minor edits to the description.

\begin{figure}[t!]

\begin{framed}
\noindent When you connect to the Internet you do so through some organization's network. For example, at home you connect to your Internet service provider's (ISP) network, while at work you connect to your employer's network. To protect your information from others on the network you can create secure connections to the websites you use (HTTPS). This is done automatically for you when you log into a website. The secure connection encrypts your Internet traffic so that no one else can view or modify your communication with the website (see Figure A).\\

\includegraphics[width=\columnwidth]{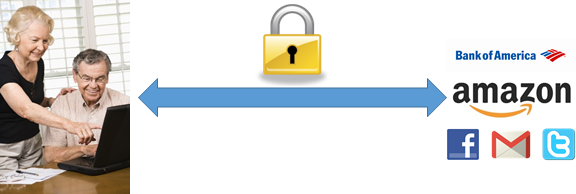}
\centerline{\textbf{Figure A}}
\vspace{.25\baselineskip}

\noindent The network you use to connect to the Internet can also be set up to use a system called a TLS proxy. TLS proxies sit in the middle of your secure connection to the websites you view (see Figure B). At the TLS proxy your Internet traffic is decrypted and the web proxy can view and modify it. Afterwards, the TLS proxy will then re-encrypt your traffic and forward it along. This is done silently and without the knowledge of you or the website you connect to.\\

\includegraphics[width=\columnwidth]{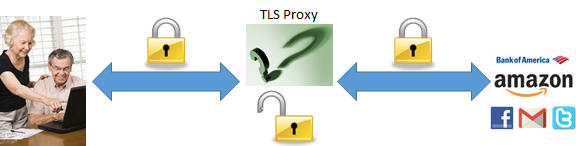}
\centerline{\textbf{Figure B}}
\vspace{.25\baselineskip}

\noindent TLS proxies can be set up by the organization that controls your Internet (for example, your ISP, school, or employer) and also by malicious attackers. TLS proxies have many different uses:\\

\vspace{-2.0\baselineskip}

\begin{center}
\resizebox{\columnwidth}{!}{
  \begin{tabular}{ll}
    {\bf Protective} & {\bf Malicious} \\
    Blocking malware and viruses & Stealing passwords \\
    Protecting company secrets & Identity theft \\
    Blocking harmful websites & Tracking government dissidents \\
    Catching malicious individuals & Spying (for example the NSA) \\
    & Censorship \\
  \end{tabular}}
\end{center}

\vspace{-1.0\baselineskip}
\end{framed}

\caption{TLS Proxy Description}
\label{fig:description}

\end{figure}

Finally, we tested this revised description using MTurk to ensure that participants felt that the description was sufficiently understandable.
Of the 80 participants in this pilot survey, nearly all participants (73; 91\%) indicated that the description of TLS proxies helped them understand what TLS proxies are and how they are used (2 participants indicated the description was not helpful (2; 3\%), with the remainder being undecided (5; 6\%)).
We also examined participant responses to free response questions and found that, as reported, most participants' answers reflected an accurate understanding of TLS proxies.
As such, we included this version of the description in both surveys, as shown in Figure~\ref{fig:description}.

\subsection{Survey Contents}
The survey begins by gathering demographic information.
It then instructs participants about TLS proxies and their use in the inspection of encrypted traffic.
Next, participants are asked to share their opinions regarding the use of TLS proxies and the inspection of encrypted traffic.
These questions survey participant opinions as to whether TLS proxies are a breach of their privacy and whether there are acceptable uses for TLS proxies.
Participants are also asked their reasoning for why TLS proxies should or should not be allowed.
Also, participants are asked which parties they are concerned about using TLS proxies and what, if any, measures should be used to regulate their use.

The survey then asks participants about how they would personally react to having a TLS proxy on a network they use to connect to the Internet.
This section includes two open-ended questions, the first asking them what concerns they might have and the second asking them how it would affect their opinion of the organization running the TLS proxy.
Finally, participants are given a chance to express any remaining comments they might have.\footnote{As shown in the Appendix, questions are grouped onto several pages. After questions on one page are answered and the user continues with the survey, they are unable to return and modify their answers.}



\subsection{Survey Development}

Before running our survey, we conducted a pilot survey using MTurk to ensure that we would get meaningful and thoughtful results.
This pilot survey was IRB approved and included 80 participants.
Based on our analysis of participants' answers in this pilot survey, it was clear that participants generally understood the description of TLS proxies presented to them, and so we proceeded to launch the full survey.
Responses from the pilot survey are not included in our results.

\subsection{Qualitative Data Analysis}
\label{sec:coding}

To better understand participants' opinions regarding TLS proxies and to avoid biasing their responses, we included several open-ended questions in the survey.
For each question, we created a codebook to categorize participant responses.
One researcher reviewed all the participant responses and created the initial codebooks.
The codebooks were then modified through discussion with the coders.

After coding was completed, all of the coders met together to discuss the data.
As part of this discussion they were encouraged to identify themes that they had seen in the data. Particular attention was paid to the themes that they felt the codebook did not adequately cover. Coders also shared responses that they felt best represented the various viewpoints expressed by participants.

In total, there were seven coders that analyzed the data.
We validated the consistency of the coders using Fleiss' Kappa \cite{fleiss1971measuring}.
Coders' agreement ranged from ``substantial agreement'' to ``almost perfect agreement'' (with kappa values ranging from .687 to 1, mean of .865 and median of .833).

\subsection{Amazon Mechanical Turk}
We used Amazon Mechanical Turk (MTurk) to recruit survey participants. 
MTurk has become an increasingly popular method for gathering participant data for usability studies and user surveys.
Buhrmester et al. found that MTurk participants are significantly more diverse than typical American College samples and that data obtained from MTurk studies is at least as reliable as those obtained via more traditional methods \cite{buhrmester2011amazon}.
Kittur et al. used MTurk participants to classify Wikipedia entries and found that that they could produce results equivalent to expert raters \cite{kittur2008crowdsourcing}.
While MTurk has known limitations, it is still a mostly reliable platform for rapidly obtaining results related to user sentiment ~\cite{ur2012does,kelley2010conducting}.

\subsection{Quality Control}
To ensure participants provided valid data, we accepted only participants that had previously completed 1,000 tasks on MTurk with an overall task approval rate of 95\% or higher.
Second, the seven coders examined participants' responses to open-ended questions in order to ensure that participants had both understood the description of TLS proxies and remained on topic. We validated the consistency of the coders' choice to exclude participants' responses using Fleiss' Kappa \cite{fleiss1971measuring} and found that coders were in perfect agreement (kappa value of 1).
During the coding process, a participant's responses were discarded if their answers were clearly spam (i.e., copying the text of a Wikipedia page), or they did not understand the questions being asked (i.e., their answers discussed HTTP proxies).
In total, we excluded 153 participants' responses (12.1\%) as spam and 60 participants (4.8\%) as misunderstandings.
The remaining 1,049 participants' responses constitute the results of our first survey.

\subsection{Demographics}

\begin{table}[htp!]
  \small
  \centering
  \rowcolors{2}{gray!25}{white}
  {\setlength{\extrarowheight}{2pt}
\begin{tabularx}{\columnwidth}{Xcc}

	\toprule
	\rowcolor{white} & \shortstack{Survey 1\\(N=1,049)} & \shortstack{Survey 2\\(N=927)} \\

	\midrule
	\multicolumn{3}{c}{Country} \\
	\midrule
	United States 	& 	86.9\%	& 94.3\% 	\\
	India 			&	11.5\%	& 5.7\%	\\
	\noalign{\global\rownum=1}Other 			& 	0.3\%	& N/A 	\\
	
	\midrule
	\multicolumn{3}{c}{Gender} \\
	\midrule
	Male 					& 	61.1\%	& 60.6\% 	\\
	Female 					&	38.6\%	& 38.9\%	\\
	\noalign{\global\rownum=1}Prefer not to answer 	&	0.3\%	& 0.4\%	\\

	\midrule
	\multicolumn{3}{c}{Age} \\
	\midrule
	18--24 years old 		& 	18.7\%	& 17.8\% 	\\
	25--34 years old 		&	47.0\% 	& 45.8\%	\\
	35--44 years old		& 	19.6\%	& 21.8\% 	\\
	45--54 years old 		&	8.6\%	& 7.9\%	\\
	55+ years old 		    & 	5.8\%	& 6.3\% 	\\
	\noalign{\global\rownum=1}Prefer not to answer 	&	0.3\%	& 0.4\%	\\

	\midrule
	\multicolumn{3}{c}{Relationship} \\
	\midrule
	Single 				& 	59.5\%	& 60.9\% 	\\
	Married 			&	35.5\%	& 35.6\%	\\
	Other 				& 	4.7\%	& 2.7\% 	\\
	\noalign{\global\rownum=1}Prefer not to answer 	        &	0.6\%	& 0.8\%	\\

	\midrule
	\multicolumn{3}{c}{Children} \\
	\midrule
	Yes 					& 	36.6\%	& 32.5\% 	\\
	No 						&	62.3\%	& 67.2\%	\\
	\noalign{\global\rownum=1}Prefer not to answer 	& 	0.9\%	& 0.3\% 	\\

	\midrule
	\multicolumn{3}{c}{Education} \\
	\midrule
	No diploma 							& 	1.0\%	& 0.6\% 	\\
	High school 						&	12.4\%	& 11.0\%	\\
	Some college or university credit 	& 	28.9\%	& 29.3\% 	\\
	College or university degree 		&	49.9\%	& 50.5\%	\\
	Post-Secondary Education 			&	7.6\%	& 8.4\%	\\
	\noalign{\global\rownum=1}Prefer Not To Answer 				& 	0.2\%	& 0.1\% 	\\

	\midrule
	\multicolumn{3}{c}{Knowledge} \\
	\midrule
	No Knowledge 				& 	4.6\%	& 2.6\% 	\\
	Somewhat Knowledgeable 	    &	35.7\% 	& 32.4\%	\\
	Mildly Knowledgeable 		& 	42.4\%	& 47.8\% 	\\
	Highly Knowledgeable 		&	14.4\% 	& 15.2\%	\\
	Expert 						& 	2.4\%	& 1.8\% 	\\
	\noalign{\global\rownum=1}Prefer Not To Answer 		& 	0.2\%	& 0.2\% 	\\

       \bottomrule
\end{tabularx}}

\caption{Participant Demographics}
\label{tab:demographics}
\end{table}

The demographics for the participants are shown in Table~\ref{tab:demographics}. 
Most participants were from the United States (87\%), with the rest primarily from India (11.5\%). Although results from a previous paper suggested that MTurk participants from India are less concerned with privacy \cite{kang2014privacy}, the results from our first survey found that they were more likely to report privacy concerns than their counterparts from the United States of America ($\chi^2[2, N=1049]=12.35, p<0.01$).

Participants were skewed towards males (61\%), and ages were centered around 25--32 (46\%).
Most participants were single (60\%) and had no children (62\%). 
Nearly all participants had completed high school, with the majority having completed some level of higher education (57\%).

Participants were asked to self-report their level of knowledge of Internet security, with most rating somewhere between somewhat knowledgeable and mildly knowledgeable (78\%).

After reading the description of TLS proxies, participants were asked whether they had prior knowledge of TLS proxies.
Most participants reported having little to no awareness of TLS proxies before the survey: unaware (66.5\%), unsure (8.1\%), aware (25.4\%).
We speculate that due to the effects of illusory superiority, the number of participants that were unaware of TLS proxies before the survey was even higher than reported \cite{glenberg1982illusion, hoorens1995self}.
Additionally, participants may have conflated knowledge of traditional web proxies with knowledge of TLS proxies.

\subsection{Limitations}
In our survey, participant demographics were slightly skewed towards a younger male population and nearly all participants were from the US and India.
Additional work could be done to replicate our results with different populations.
Cross-cultural, international surveys would be especially interesting, but these should be conducted by researchers that can engage participants in their native language and have an understanding of participants' cultural perceptions.

As shown in prior work, participants' reported security preferences and desires do not always align with their actual behaviors~\cite{woodruff2014would}.
Often users will report being more privacy minded than they are in practice.
Interestingly, in our survey participants indicated a high level of acceptance for TLS proxies, which could suggest that real-world acceptance of TLS proxies is even higher than we measured.
On the other hand, many participants reported wanting to have their consent obtained, or at least be notified of, the inspection of encrypted traffic; in practice, it is possible that fewer participants would actually be interested in being notified.

Finally, while we spent considerable effort to craft a fair and unbiased description of TLS proxies and the inspection of encrypted traffic, there is still the possibility that it had a significant effect on some participants' responses.
For example, in the real world, users often learn about security issues from the news, which is often sensational and biased.
In contrast, our description strove for neutrality, and as such may have led to users taking a more rational view of the inspection of encrypted traffic than would occur in the wild.
While we chose to accept these limitations in order to obtain opinions from as many participants as possible, an open avenue for future research is to find a way to gather equally widespread opinions in a way that has fewer limitations.

\section{First Survey -- Results}

\begin{figure*}[t]
  \centering
  \includegraphics[width=\textwidth]{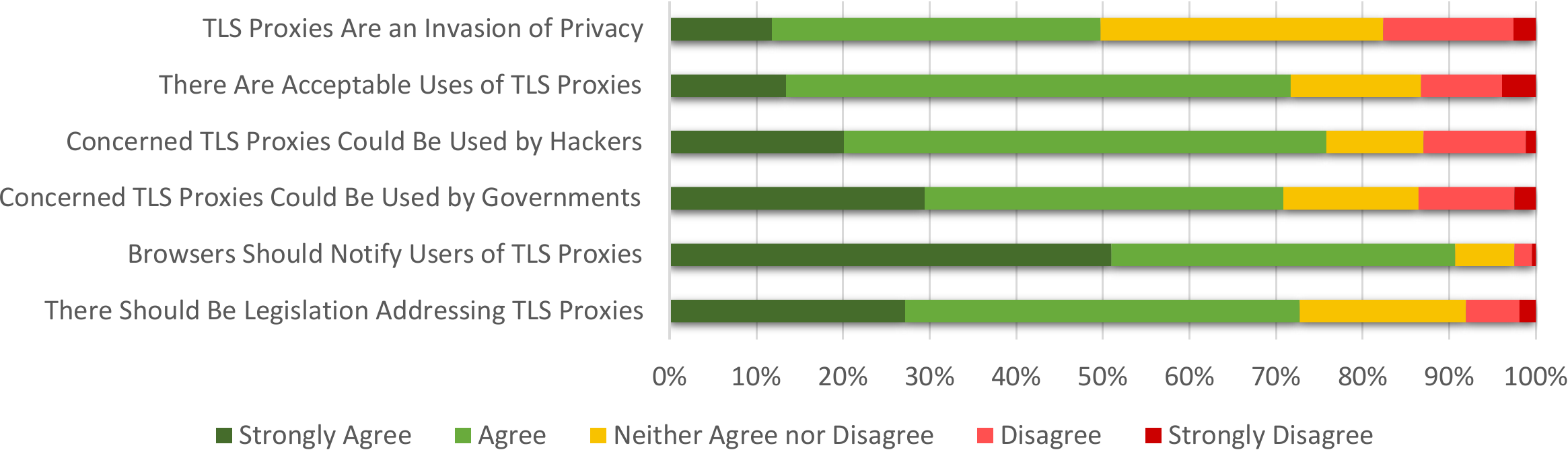}
  \caption{Participant Attitudes Toward TLS Proxies (N=1,049)}
  \label{fig:questionsLikert}
\end{figure*}


In this section we discuss the results of our survey in three areas:
acceptable uses for TLS proxies, 
general concerns toward their use, and
the reaction participants would have if they discovered a network they
use employed a TLS proxy. 


\subsection{Acceptable Uses of TLS Proxies}
\label{sec:uses}

Figure~\ref{fig:questionsLikert} shows participant attitudes toward proxies.
A somewhat surprising result is that participants largely (752; 71.7\%) felt that there were acceptable uses for TLS proxies.
This feeling prevailed even though nearly half of the participants (522; 49.8\%) indicated that TLS proxies are an invasion of privacy, and only one-eighth of participants (185; 17.6\%) felt they presented no invasion of privacy.
There is a strong correlation between thinking TLS proxies were an invasion of privacy and believing that there were not acceptable uses for them ($\chi^2[4, N=1049]=141.50, p<0.001$).
Nevertheless, over a quarter of participants (297; 28.0\%) felt that TLS proxies were an invasion of privacy, but still had acceptable uses.

To better understand what uses might be acceptable, we asked
participants who felt there were acceptable uses to enumerate
those uses in an open-ended question.  
The results from our coded responses are shown in the
top part of Table~\ref{tab:questionsOpinions}. The acceptable uses
are largely concentrated on three use cases:

\begin{table}
  \small
  \centering
  \rowcolors{2}{gray!25}{white}
  {\setlength{\extrarowheight}{2pt}
\begin{tabularx}{\columnwidth}{Xc}	
    \toprule
	\rowcolor{white}Opinion & Participants \\
    
    \midrule
    \multicolumn{2}{c}{Acceptable Uses} \\
    \midrule
    
    Protect organizations & 51.4\% (n=539) \\
    Protect individuals & 34.8\% (n=365) \\
    Law enforcement and surveillance & 8.9\% (n=93) \\
    Censor content & 7.1\% (n=75) \\
    Never censor content & 3.1\% (n=32) \\
    \noalign{\global\rownum=1} Acceptable at work, not at home & 2.9\% (n=30) \\
    
	\midrule
    \multicolumn{2}{c}{Concerns} \\
    \midrule
    
    Hackers and spying & 60.5\% (n=635) \\
    Privacy and identity theft & 55.4\% (n=581) \\
    \noalign{\global\rownum=1} Done without knowledge or consent & 13.2\% (n=138) \\
    
    \midrule
    \multicolumn{2}{c}{Reactions} \\
    \midrule
    
    Negative & 60.8\% (n=638) \\
    Positive & 5.0\% (n=52) \\
    Depends & 34.2\% (n=359) \\
    Suspicious & 25.8\% (n=271) \\
    Discontinue use & 17.2\% (n=180) \\  
    \noalign{\global\rownum=1} Change behavior (besides discontinue) & 6.2\% (n=65) \\
	
	\bottomrule
\end{tabularx}}

\caption{Qualitative Response Categorization (N=1,049)}
\label{tab:questionsOpinions}
\end{table}

\begin{enumerate}

\item {\bf Protecting organizations (493; 65.6\%).}
Many participants felt that organizations (e.g., businesses, government agencies, schools, libraries) had a right to protect their own intellectual property and security.
This included protecting the company from viruses and hackers, filtering inappropriate or potentially malicious websites, and preventing the leak of sensitive information.
Participants mentioned that since these organizations provide the Internet for their employees or constituents they had a right to use TLS proxies on their own networks.

\item {\bf Protecting individuals (339; 45.1\%).}
Participants saw value in businesses using TLS proxies to protect their customers.
This protection came in one of two forms:

\begin{itemize}
\item {\bf Direct.} Antivirus applications and firewalls could use TLS proxies to filter malware and viruses. Similarly, ISPs could use TLS proxies to detect and prevent phishing attackers and block other inappropriate or malicious websites.

\item {\bf Indirect.} Participants recognized that they have a significant amount of private information stored externally on the web (e.g., at Amazon or Google).
In order to protect this data, participants hoped that the companies storing their private data would employ TLS proxies internally to ensure the safety of the customer's data.
\end{itemize}



\item {\bf Law enforcement and surveillance (65; 8.6\%).}
Nearly a tenth of participants expressed that law enforcement agencies should also be allowed to use TLS proxies.
This includes use by local or federal agencies to track criminal or terrorist activity.
Several participants also expressed that while this was a legitimate use it should only be done with a valid warrant or if there was an imminent threat to national security.

\end{enumerate}

\subsection{Concerns}
\label{sec:concerns}

Even though many participants in the first survey saw acceptable uses for TLS proxies, they were not without concerns or reservations. Based on our coding, we grouped these concerns into the categories shown in the middle part of Table~\ref{tab:questionsOpinions}.
Three-quarters of the participants (795; 75.8\%) mentioned they worried about hackers and nearly as many were concerned about the possibility for governmental spying (743; 70.9\%).
There was also a strong correlation between the concern that hackers could use TLS proxies and that the government could use them ($\chi^2[4, N=1049]=194.57, p<0.001$).

The most visceral concerns were related to the breach of privacy.
One of the open response questions asked participants to list what possible concerns they had regarding the use of TLS proxies.
Over half of participants  (581; 55.4\%) mentioned they were concerned with a loss of privacy and personal information.
Nearly a tenth of participants (104; 9.91\%) mentioned having their identity stolen, and even more participants had answers that addressed the issue of identity theft generally.

A non-negligible number of the participants freely shared that either they, a family member, or other acquaintance had been the victim of account compromise.
Similar to the finding of Shay et al. \cite{shay2014my} this was a traumatic experience and it left participants especially concerned that TLS proxies could be used to perpetrate identity theft. R208 shared,
\begin{quote}
{\it ``A major concern that I would have would be the security of my personal and financial information. I have many friends who have been victims of identity theft and fraud, and would hate to have to go through what they did.''}
\end{quote}

Participants were also concerned that TLS proxies could be used without their knowledge.
One-eighth of participants (138; 13.2\%) mentioned in the open response question that they were concerned with privacy.
Furthermore, when directly asked about notification, an overwhelming majority of participants (951; 90.7\%) asserted they wanted to be notified by their browsers of the presence of TLS proxies.
Similarly, participants largely (942; 89.8\%) felt that there should be legislation concerning TLS proxies. Most (782; 74.5\%) wanted legislation to require notification, and nearly as many (701; 66.8\%) wanted legislation to require consent.


\subsection{Reactions}
\label{sec:reactions}

Participants in the first survey had varied responses on how they would react to learning that they currently use a network that employs TLS proxies.
Based on our coding, we grouped these concerns into the categories shown in the bottom part of Table~\ref{tab:questionsOpinions}.
Over half of participants (638; 60.8\%) mentioned that it would negatively affect their opinion of the owner of that network.
For example, R77 stated,
\begin{quote}
{\it ``I would be angry and would feel that organization violated my trust. I would wonder what information that organization had been collecting on me and what they planned to do with it. If it was my employer, I also would think that organization did not trust me and would consider working somewhere else.''}
\end{quote}

Still, a third of participants (359; 34.2\%) said that their reaction would depend on who the owner of the network was and how they were using the proxy.
For example, if the owner of the network was their employer they would not have a negative reaction, but if it was their ISP or government they would be very unhappy.
Participants also mentioned that their approval would rest on whether or not any personal information was collected and/or sold and whether their consent had first been obtained.
R960 explained,
\begin{quote}
{\it ``It would be on a case by case basis. I can see some instances where it would be understandable, but if it was going on without my consent, I would be wary of dealing with them in the future.''}
\end{quote}

Participants also mentioned ways in which their behavior would change if they learned a network was employing a TLS proxy.
A quarter of participants (271; 25.8\%) said that it would make them suspicious of the owner of that network.
A quarter of participants (245; 23.4\%) also mentioned that they would change their behavior on that network.
For some participants (180; 17.2\%) this included discontinuing use of the network and its services, while others (65; 6.2\%) mentioned they would change the content they looked at on the Internet or be more careful about entering personal information, including but not limited to e-commerce transactions.
At the extreme, some participants mentioned they would quit their job if they found that their employer's network used a TLS proxy.
For example, R127 expressed,
\begin{quote}
{\it ``If my employers were secretly spying on my private data, I would sue them if legally possible, and quit the job regardless.''}
\end{quote}

\newpage
\subsection{Personas}

As our research group discussed the answers to open response questions in the first survey, it became clear that the participants could generally be classified into one of four personas: {\em pragmatic}, {\em privacy fundamentalist}, {\em jaded}, and {\em unconcerned}.
After recognizing this, two members of the research group re-evaluated 90 participant responses and categorized participants into one of these four personas.
The Fleiss' Kappa for this classification was 1 (i.e., perfect agreement).
One researcher then classified the rest of the responses.
The breakdown of participants into these categories is given in Table~\ref{tab:persona}.\footnote{There were ten participants whose answers were vague enough that we did not feel comfortable classifying them as any of the personas.}


Even though three of these personas have similar names to personas formulated by Westin~\cite{westin1991harris}, our categories are in no way based on the research of Westin.
Instead, our methodology for creating personas more closely relates to that of Woodruff et al.~\cite{woodruff2014would}, i.e., analyzing how participants indicate they would act in various privacy-related situations in order to determine their persona.
Moreover, we do not intend these personas to be a definitive list of privacy personas, but rather view them as a helpful way to identify trends within our data.


\begin{table}
  \small
  \centering
  \rowcolors{2}{gray!25}{white}
  {\setlength{\extrarowheight}{2pt}
\begin{tabularx}{\columnwidth}{Xcc}
	
    \toprule
	\noalign{\global\rownum=2}\rowcolor{white}Persona & Number & Percent \\
    \midrule

	Pragmatic majority & 802 & 76.5\% \\
	Privacy fundamentalist & 178 & 17.0\% \\
	Jaded & 48 & 4.6\% \\
	Unconcerned & 11 & 1.0\% \\
    Unclassified & 10 & 1.0\% \\

	\bottomrule
\end{tabularx}}

\caption{Participant Persona Categorization (N=1,049)}
\label{tab:persona}
\end{table}

\subsubsection{Pragmatic Majority, N=802}
The pragmatic majority weighed consumer benefits and
protections of public safety against costs of intrusive practices,
believed that organizations should earn the public's trust, and wanted
to have the opportunity to opt-out of intrusive practices.
This group was strongly correlated with being more likely to feel that there were acceptable uses for TLS proxies ($\chi^2[4, N=1028]=230.48,$ $p<0.001$).
R93 stated,
\begin{quote}
{\it ``I think it is perfectly acceptable for organizations (companies, schools, libraries, etc.) to use TLS proxies because it protects their computers.  It keeps hackers from getting to sensitive or confidential information of the organization.  In addition, it blocks harmful viruses that can cause a lot of damage and expense in repair.  It can also keep individuals from accessing websites (employees from playing online games or minors from accessing pornography).  It is perfectly reasonable for companies to employee[sic] this device for these purposes when an individual is using their computer.  We should not expect privacy when we are using someone else's computer.''}
\end{quote}

Though the pragmatic majority all weighed consumer benefits versus intrusive practices, they were not uniform in their conclusions about where and how TLS proxies should be used.
Some recognized the right of employers to use them, while others believed they should only be
allowed in narrow cases such as with a warrant.

\subsubsection{Privacy Fundamentalist, N=178}

The privacy fundamentalist was generally distrustful of
organizations that ask for personal information, in favor of
legislation enhancing privacy, and chose privacy controls over
consumer benefits when a trade-off existed between the two.
These participants were strongly correlated with being more likely to feel TLS proxies were an invasion of privacy ($\chi^2[4, N=1028]=114.81, p<0.001$).
These participants were also more likely to support legislation of TLS proxies ($\chi^2[2, N=1028]=14.40, p<0.001$).

The defining feature of the privacy fundamentalist was that they viewed privacy as so important that it could not be traded for any benefit, no matter how great. As emphatically stated by R1119, {\it ``I believe privacy is sacrosanct and one could argue that it's a Constitutional right.''}

They were also likely to relate the use of TLS proxies to more traditional methods of surveillance such as wiretapping and intercepting mail.

\subsubsection{Jaded, N=48}

Jaded individuals were aware that violations of privacy happen regularly,
believed that governments conduct surveillance on the general
public, and had lost hope that they can have privacy online.
These participants felt that ``the system'' was rigged to remove any real chance of them having privacy. For example, R713 expressed,
\begin{quote}
{\it ``I know that it is my choice to use the internet; however, since I live in a remote area with no transportation to the nearest city (30 miles away) I am `stuck' working and banking and doing business on the internet. I feel it is unfair to be made to choose between being `safe' and having privacy freedom. I am especially disgusted by our government's spying behaviors and the rhetoric about it being necessary for national defense.''} 
\end{quote}

Likewise, when asked about concerns regarding the use of TLS proxies, R831 shared,
\begin{quote}
{\it ``None. The government (via the NSA) is already reading everything we do and share online. So no surprises there.''}
\end{quote}
Other jaded participants felt they had no choice in the matter because in the United States Internet service providers often have a monopoly.


\subsubsection{Unconcerned, N=11}

Unconcerned participants were generally trustful of organizations that ask
for personal information, willing to sacrifice personal privacy to
obtain consumer benefits, and not in favor of legislation to protect or enhance privacy.
In our survey, we found very few unconcerned participants (1\%).
It is possible that the recent news regarding widespread government surveillance caused participants to be more privacy aware and sensitive.
In addition, our use of qualitative data to classify participants allowed us to recognize that participants were part of the pragmatic majority even when their Likert responses might seem to indicate otherwise.

\section{Second Survey -- Methodology}
Our first survey revealed that participants' opinions related to TLS proxies were closely tied to the situation in which TLS proxies were being used.
To better clarify user feelings in this area, we formulated a second survey in which we ask participants about a series of specific scenarios where inspection of encrypted traffic could be used.
This second survey serves to give quantitative backing to the qualitative data gathered in the first survey.

We collected data for our second survey on Tuesday, February 24, 2015 between 11:02 AM and 1:06 PM (PST). Each participant could take the survey once and received \$1 USD as compensation upon completing the survey.
The survey begins exactly as the first survey by gathering demographic information and
then instructing participants about TLS proxies and their uses, both benevolent and malicious.
Participants are then asked their opinions regarding the use of TLS proxies in various circumstances.
In total 1,005 people completed the online survey.
The survey was also approved by our Institutional Review Board and is contained in Appendix~\ref{appx:study-two}.

\subsection{Survey Description}
The first portion of the second survey includes the same description of TLS proxies as the first one. It then asks several questions repeated from the first survey: whether TLS proxies are an invasion of privacy and whether there are acceptable uses for TLS proxies.

The main portion of this survey asks participants their opinion
regarding different situations where TLS proxies may be used to inspect encrypted traffic, such as by an employer, at a school, or a caf\'e with free WiFi. The full list of scenarios is given in Figure~\ref{fig:scenarios}. For each situation, participants are asked whether the organization should be
allowed to run a TLS proxy, with responses taken from (1) {\em No}, (2) {\em   Only if I consent}, (3) {\em Only if I am notified (consent not required)}, (4) {\em Yes (neither notification nor consent required)}, or (5) {\em Unsure}. To choose the situations, we used responses from open-ended questions in the first survey, along with suggestions from our research team to fill out the list.
Finally, we had a single open-ended question where participants could share any opinions they still had remaining at the end of the survey.

We note that this survey had the same limitations as our first survey.

\begin{figure*}[t]
  \centering
  \includegraphics[width=\textwidth]{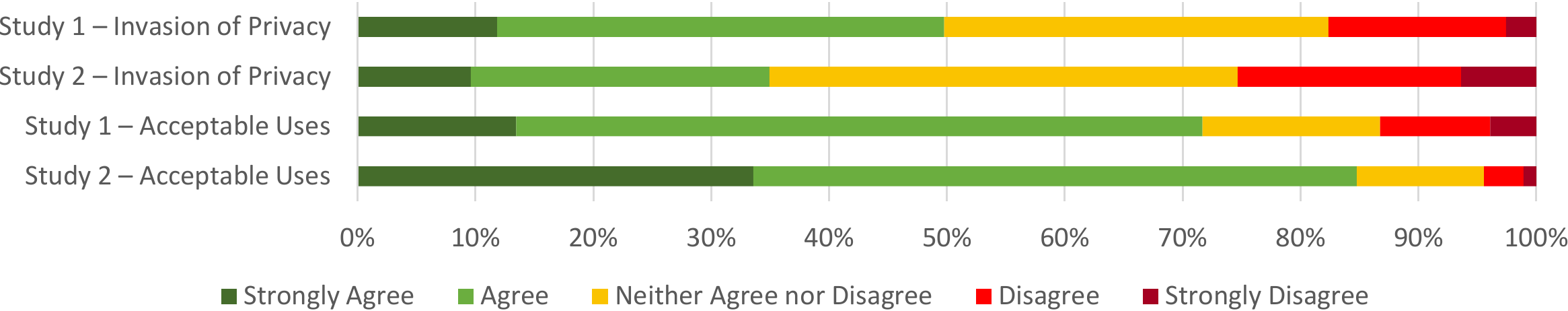}
  \caption{Participant Attitudes Toward TLS Proxies (Survey 1 -- N=1,049, Survey 2 -- N=927)}
  \label{fig:questionsLikert2}
\end{figure*}

\subsection{Quality Control}
To ensure participants provided valid data, we accepted only participants that had previously completed 1,000 tasks on MTurk with an overall task approval rate of 95\% or higher.
Second, we limited participants to the United States and India.
This was done because with the first survey coders struggled to understand answers to free response questions from outside the United States and India.\footnote{Moreover, these represent a small enough portion of participants that their responses had no significant effect on the data.}
Third, we looked at the single open-ended question to determine if participants had entered spam (e.g., copied an answer from Wikipedia).
Finally, we used two validation questions in the second survey because there were not enough open responses to always distinguish spam entries.

In total, we excluded 78 participant's responses (7.8\%).
The remaining 927 participant's responses constitute the results of our second survey.
 
\subsection{Demographics}

The demographics for the participants were summarized earlier in Table~\ref{tab:demographics}.
There were no significant differences in the demographics of the first and second surveys.

\begin{table}
  \small
  \centering
  \rowcolors{2}{gray!25}{white}
  {\setlength{\extrarowheight}{2pt}
  \begin{tabularx}{\columnwidth}{Xcc}

 	\toprule
 	\rowcolor{white} & \shortstack{Survey 1\\(N=1,049)} & \shortstack{Survey 2\\(N=927)} \\

 	\midrule
 	\multicolumn{3}{c}{Prior Knowledge of TLS Proxies} \\
 	\midrule
 	Strongly Agree 				& 	4.1\%	& 8.4\% 	\\
 	Agree 	        			&	21.3\% 	& 27.9\%	\\
 	Neither Agree nor Disagree 	& 	8.1\%	& 13.2\% 	\\
 	Disagree 					&	48.1\% 	& 34.3\%	\\
 	Strongly Disagree 			& 	18.4\%	& 16.2\% 	\\

 	\bottomrule

  \end{tabularx}}

  \caption{Participants' Knowledge of TLS Proxies}
  \label{tab:knowledge}
\end{table}


\section{Second Survey -- Results}

In this section we discuss results from our second survey.
First we compare results from the three questions that were the same between both surveys.
We then discuss the quantitative data regarding participants' opinions regarding different deployment scenarios for TLS proxies.

\subsection{Comparison}

In both surveys, after reading the description of TLS proxies, participants were asked whether they had prior knowledge of TLS proxies.
These are shown in Table~\ref{tab:knowledge}.
In the first survey, most participants reported having little to no awareness of TLS proxies before the survey: aware (25.4\%), unsure (8.1\%), unaware (66.5\%).
In the second survey, more participants reported being aware of proxies beforehand (the difference is statistically significant, $\chi^2[4, N=1976]=60.003, p<0.001$), though over half still reported having little to no awareness of TLS proxies before the survey: unaware (50.5\%), unsure (13.2\%), aware (36.3\%).\footnote{As before, we speculate that due to the effects of illusory superiority, the number of participants that were unaware of TLS proxies before the survey was even higher than reported \cite{glenberg1982illusion, hoorens1995self}.}

We also compared responses relating to whether participants in both surveys felt that TLS proxies were an invasion of privacy, and whether TLS proxies had acceptable uses (see Figure~\ref{fig:questionsLikert2}).
Participants in the second survey were less likely to view TLS proxies as an invasion of privacy (first survey -- 50\%, second survey -- 35\%), with the difference being statistically significant ($\chi^2[4, N=1976]=54.228, p<0.001$).
Similarly, participants in the second survey were also more likely to feel that there were acceptable uses for TLS proxies (first survey -- 72\%, second survey  -- 85\%), with this difference also being statistically significant ($\chi^2[4, N=1976]=140.654, p<0.001$).

It is important to note that in both surveys, after participants answered each group of questions (see Appendix) participants were unable to return to earlier groups of questions and alter their answers.
As such, the above reported differences are not due to differences in the survey, as up to this point the surveys were identical.

\subsection{Scenarios}
\label{sec:scenarios}

\begin{figure*}[t]
  \centering
  \includegraphics[width=\textwidth]{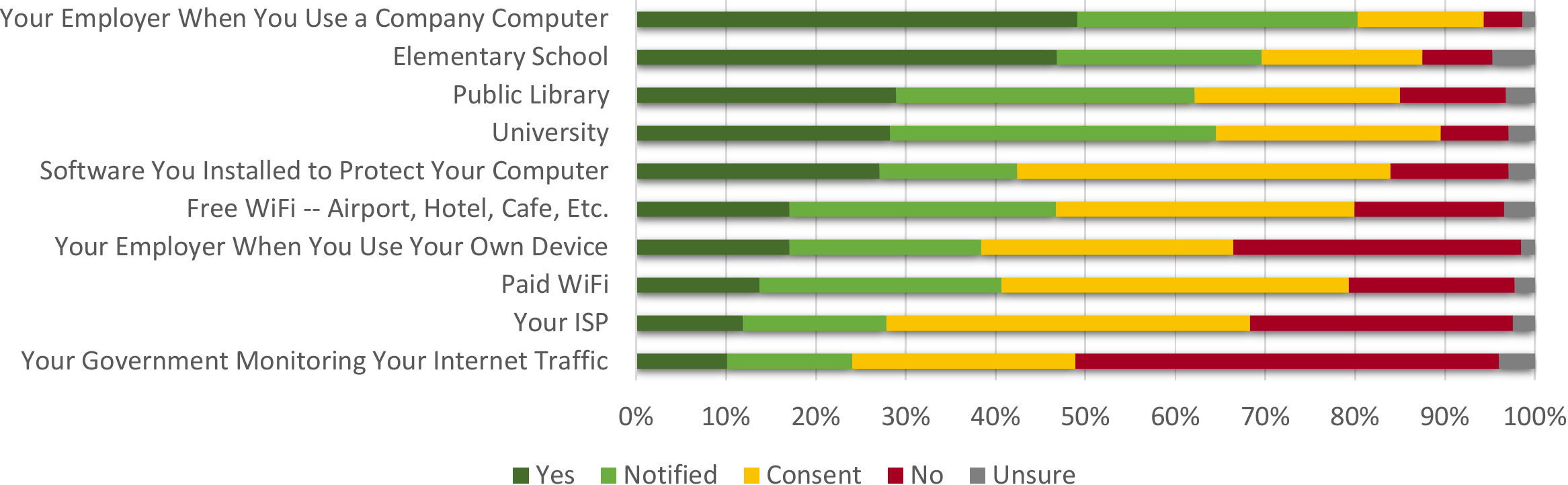}
  \caption{Participant Responses on Scenarios---Should the Organization Be Allowed To Run a TLS Proxy? (N=927)}
\label{fig:scenarios}
\end{figure*}

We asked participants regarding their opinions towards the inspection of encrypted traffic in specific scenarios.
For each scenario, participants indicate whether they were comfortable with the traffic being intercepted (``Yes''), whether they wanted to be notified (``Notified''), whether they wanted their consent to be obtained (``Consent''), or whether they were uncomfortable with it.
The results for these questions are summarized in Figure~\ref{fig:scenarios}.

Participants in our second survey are generally willing to accept the use of TLS proxies in most situations, with acceptance ranging from 65\% to 90\% of participants, when summing together those who accept it, those who desire notification, and those who desire both notification and consent.
For both employers (when you use your own computer) and elementary schools, the support for using TLS proxies without notification or consent from users is surprisingly strong (455; 49.1\% and 434; 46.8\%).
This may be due to a belief in employer rights in the first case and a desire to protect children in the second case.
In both cases there is still strong support for either notification or consent (419; 45.2\% and 377; 40.7\%).

The strongest objections to any kind of TLS proxy are for government monitoring (437; 47.1\%), using your own device at work (297; 32.0\%), or using your own ISP (271; 29.2\%). Note these latter two map to situations where 
the user has paid for the device or for network access. Users have stronger objections to TLS proxies when they pay for network access through a home ISP than when they pay for WiFi when they are away from home.

When examining the differences among opinions for notification versus consent, we see that the preference for consent is higher for personal firewalls (software you installed to protect your computer), your ISP, free WiFi, paid WiFi, and using your own device at work. The preference is higher for notification for a public library, university, elementary school, and using a company computer at work.
This seems to be a clear split that favors consent in cases where the user feels in control versus notification when an organization is in control.
The strongest support for consent is with a personal firewall (385; 41.5\%), your ISP (375; 40.5\%), and paid WiFi (358; 38.6\%).


\section{Discussion}
In this section we discuss interesting themes we saw as we analyzed participants' responses to the open-ended questions.

\subsection{Informed Participants}

Most of the participants showed a high level of engagement in the survey.
At the end of the survey when asked if they had any additional comments, a large number of participants mentioned that they were thankful that we had informed them of this information.
Some even asked where they could get more information on the topic of TLS proxies.
Additionally, we were impressed with the in-depth analysis of trade-offs that many users shared, which often went far beyond the scope of any information provided to them in the survey.

Participants clearly understood that there were trade-offs involved with the use of TLS proxies and the inspection of encrypted traffic, weighing the benevolent uses for schools or workplaces and the danger of misuse by insiders or by hackers. As they struggled with this trade-off, participant responses indicated confusion, doubt, worry, equivocation, and reasoned conclusions.
Confusion regarding how to resolve the conflict was evident when participants labeled it a ``grey area.'' R988 considered both good and bad uses and worried, 
{\it ``How are you supposed to know which is happening?''}

Some participants weighed the trade-offs and resolved the dilemma by deciding that proxies should only be used by consent.
For example, R827 expressed:

\begin{quote}
{\it ``I believe that TLS proxies are an invasion of privacy, as is
anything that monitors my internet usage without my
permission. However if you are using someone else's (like a
company's) network, they have every right to make the rules of
use... This is one of those doubled-edged swords -- it
can be used for your good and security and it can be used to harm
and spy on you. Because of the distinct possibility of lost privacy,
this type of proxy should [not be] used, except by your agreement, not
by anyone else.''}
\end{quote}
Others wanted companies or schools to be able to use TLS proxies for security
purposes, but also wanted to prevent them from being used for government surveillance or by hackers.  
Still others felt TLS proxies should {\em only} be used by the government to catch terrorists or criminals.

Similarly, of the participants who were against the use of TLS proxies, the reasons for opposing TLS proxies were not amorphous, but concrete and rational.
For example, R666 stated:

\begin{quote}
{\it ``I think TLS proxies don't sound very safe because it sounds like
an invasion of privacy. I don't think organizations should be able
to decrypt your internet traffic and modify it and re-encrypt
it. Perhaps they are just trying to protect against viruses and the
like but it doesn't sound safe for the person using the
internet. What if this technology was misused? Someone could get
[h]old of your financial information for example. It sounds to[o]
risky. I wouldn't want to buy something online and risk someone
having access to my credit card number.''}
\end{quote}

\subsection{Notification and Consent}

Numerous participants expressed a desire for notification and consent
when TLS proxies were being used on a network. A typical response as
given by R413 was,
\begin{quote}
{\it ``Well for some things it would be
  understandable, I'd just like to be informed so I know the risk I'm
  taking.'' }
\end{quote}  
  
R313 expressed,
\begin{quote}
{\it ``If I encrypt something no one has
  the right to unencrypt it unless I give them the right to - simple
  as that.''}
\end{quote}
  
Participants expressed extreme distrust for those who would use TLS
proxies without informing users, going so far as to say they {\em ``would
hate them,''} {\em ``would wonder what they are looking for,''} and {\em ``would
assume they were up to no good.''}

Others stated they would change their behavior if notified about a
proxy, such as avoiding commercial transactions, using a VPN to
circumvent a proxy, or self-censorship of their Google searches and other online communication.

\subsection{Jaded Participants}
We were surprised to find that 4.5\% of participants were ``jaded'' towards the current state of privacy online.
They felt that currently it is largely impossible to have any expectation of privacy or security.
Many felt that the government was already spying on the population at large, and that even without TLS proxies the government could find a way to gain access to their private information.
Others felt that even if they discovered that their traffic was being intercepted, they would have no recourse as their access to the Internet is controlled by a monopoly.

We find this group concerning, as this is not a group of individuals unconcerned with security and privacy.
Rather they are a group that still cares about privacy, but has lost all hope that they can actually achieve digital privacy.
This is a troubling trend, as such individuals are unlikely to adopt solutions that could actually benefit them.
As such, work needs to be done to determine how this type of user's trust can be regained.

\subsection{Changing Opinions}
\label{sec:discuss-change}
Between our two surveys, we noticed differences in the way participants viewed TLS proxies.
This demonstrates that users' perceptions towards security and privacy are not static.
As such, it is important that work such as this be done on a regular basis, helping the security community stay abreast of current opinions and attitudes.

One interesting difference is that in the second survey fewer participants viewed inspection of encrypted traffic as an invasion of privacy, and more participants felt that there were acceptable uses for this practice.
One possible explanation for this difference
is that news stories have been discussing how encryption and other privacy preserving technologies could be used by terrorist organizations.
Still, additional research is needed to better understand this shift in attitudes towards security and privacy.

\section{Conclusion}

This paper presents the first survey of general (i.e., non-expert) user attitudes toward TLS proxies. 
Responses indicate that participants hold nuanced opinions on security and privacy trade-offs, with most recognizing legitimate uses for the proxies, but concerned about threats from hackers or government surveillance.  
A significant concern about malicious uses of TLS inspection is identity theft, and many would react negatively and some would change their behavior if they discovered inspection occurring without their knowledge.
We also find that a small but significant number of participants are jaded by the current state of affairs and have lost any expectation of privacy.

User attitudes toward TLS proxies provide an important data point along the spectrum of discussion that is currently taking place regarding who should have access to encrypted information. 
The results of our survey demonstrate that participants were generally aware of the trade-offs between privacy and security, and that most participants were willing to sacrifice some privacy for additional security.
Nevertheless, participants strongly supported notification and consent for when encrypted traffic is being inspected.
%

\section{Acknowledgment}
The authors thank Rich Shay for providing feedback on the wording of questions in our first survey.
We also thank Alexander Lemon, JJ Lowe, Brent Roberts, and Justin Wu for help with coding the data. 
Finally, we thank the anonymous reviewers for their helpful comments.

\bibliographystyle{abbrv}
\bibliography{paper}

\clearpage
\appendix

\section{First Survey}
\label{appx:study-one}
\vspace{\baselineskip}

\subsection{Page 1}

\noindent We are conducting an academic research survey about public opinions on Internet security. The survey will take approximately 5 minutes.\\

\noindent We will not collect any personally identifying information. If you do not complete the survey we will not store any of your responses. If you have any questions or concerns about the information collected, please contact us at [email redacted].

\subsection{Page 2}

\noindent \textbf{What is your gender?}
\textit{
\begin{itemize}[label={$\circ$},leftmargin=10pt,noitemsep,nolistsep]
\item Male
\item Female
\item I prefer not to answer\\
\end{itemize}
}

\noindent \textbf{What is your age?}
\textit{
\begin{itemize}[label=$\circ$,leftmargin=10pt,noitemsep,nolistsep]
\item 18 -- 24 years old
\item 25 -- 34 years old
\item 35 -- 44 years old
\item 45 -- 54 years old
\item 55 years or older
\item I prefer not to answer\\
\end{itemize}
}

\noindent \textbf{What is the highest degree or level of school you have completed?}
\textit{
\begin{itemize}[label=$\circ$,leftmargin=10pt,noitemsep,nolistsep]
\item Some school, no high school diploma
\item High school graduate, diploma or the equivalent (for example: GED)
\item Some college or university credit, no degree
\item College or university degree
\item Post-secondary education
\item I prefer not to answer\\
\end{itemize}
}

\noindent \textbf{What is your marital status?}
\textit{
\begin{itemize}[label=$\circ$,leftmargin=10pt,noitemsep,nolistsep]
\item Married
\item Single
\item Other 
\item I prefer not to answer\\
\end{itemize}
}

\noindent \textbf{Do you have children?}
\textit{
\begin{itemize}[label=$\circ$,leftmargin=10pt,noitemsep,nolistsep]
\item Yes
\item No
\item I prefer not to answer\\
\end{itemize}
}

\noindent \textbf{In which country do you reside?}

\subsection{Page 3}

\noindent \textbf{Where are taking this survey?}
\textit{
\begin{itemize}[label=$\circ$,leftmargin=10pt,noitemsep,nolistsep]
\item Home
\item Work
\item School
\item Library
\item Retail (coffee shop, internet cafe, etc.)
\item Other
\item I prefer not to answer\\
\end{itemize}
}

\noindent \textbf{What type of Internet connection are you using?}
\textit{
\begin{itemize}[label=$\circ$,leftmargin=10pt,noitemsep,nolistsep]
\item Wired
\item WiFi
\item Cellular (3G, 4G, etc.)
\item Other
\item I don't know
\item I prefer not to answer\\
\end{itemize}
}

\noindent \textbf{How knowledgeable are you about Internet security?}
\textit{
\begin{itemize}[label=$\circ$,leftmargin=10pt,noitemsep,nolistsep]
\item Expert
\item Highly knowledgeable
\item Mildly knowledgeable
\item Somewhat knowledgeable
\item No Knowledge
\item I prefer not to answer\\
\end{itemize}
}

\noindent \textbf{When connecting to a website securely, for example when doing online shopping or banking, who should be able to see the contents of your Internet traffic? (Choose all that apply)}
\textit{
\begin{itemize}[label=$\circ$,leftmargin=10pt,noitemsep,nolistsep]
\item Me
\item My Internet provider
\item The website
\item Malicious individuals
\item Everyone
\end{itemize}
}

\subsection{Page 4}
\label{appx:description}

\vspace{\baselineskip}

\noindent \texttt{Strongly Disagree, Disagree, Neither Agree nor Disagree, Agree, Strongly Agree}
\textit{
\begin{itemize}[label=$\circ$,leftmargin=10pt,noitemsep,nolistsep]
\item The above description of TLS proxies helped me to clearly understand what TLS proxies are and how they are used.
\end{itemize}
}

\subsection{Page 5}

\noindent \texttt{Strongly Disagree, Disagree, Neither Agree nor Disagree, Agree, Strongly Agree}
\textit{
\begin{itemize}[label=$\circ$,leftmargin=10pt,noitemsep,nolistsep]
\item Prior to taking this survey, I was aware that organizations were using TLS proxies.
\item TLS proxies are an invasion of privacy.
\item There are acceptable uses for TLS proxies.\\
\end{itemize}}

\noindent \texttt{Only seen if selected "Agree"  or "Strongly Agree" to acceptable uses for TLS proxies.}\\
\noindent \textbf{Please explain which organizations should be allowed to use TLS proxies and for what purpose.} (only shown on an Agree or Strongly Agree answer from above)\\

\noindent \texttt{Only seen if selected "Disagree"  or "Strongly Disagree" to acceptable uses for TLS proxies.}\\
\noindent \textbf{Please explain why TLS proxies should never be allowed.}

\subsection{Page 6}

\noindent \texttt{Strongly Disagree, Disagree, Neither Agree nor Disagree, Agree, Strongly Agree}
\textit{
\begin{itemize}[label=$\circ$,leftmargin=10pt,noitemsep,nolistsep]
\item I am concerned that TLS proxies could be used by hackers to compromise my Internet security.
\item I am concerned that TLS proxies could be used by the government to collect my personal information.
\item Browsers should notify users if there is a TLS proxy intercepting and decrypting their Internet traffic.
\item There should be legislation that addresses TLS proxies.\\
\end{itemize}
}

\noindent \texttt{Only seen if selected "Agree"  or "Strongly Agree" to legislation that addresses proxies.}\\
\noindent \textbf{What should legislation that addresses TLS proxies do? (Choose all that apply)}
\textit{
\begin{itemize}[label=$\circ$,leftmargin=10pt,noitemsep,nolistsep]
\item Prevent their use
\item Require organizations to obtain consent before using a TLS proxy
\item Require organizations to inform users when a TLS proxy is being used
\item I don't believe that legislation is required
\item Other
\end{itemize}
}

\subsection{Page 7}

\noindent The following statements and questions are about how you would personally react to having a TLS proxy on a network you use to connect to the Internet.\\

\noindent \texttt{Strongly Disagree, Disagree, Neither Agree nor Disagree, Agree, Strongly Agree}
\textit{
\begin{itemize}[label=$\circ$,leftmargin=10pt,noitemsep,nolistsep]
\item I believe TLS proxies are in use on a network I use to connect to the Internet.\\
\end{itemize}
}

\noindent \textbf{Please explain what concerns you have about a TLS proxy being used on a network you personally use to connect to the Internet.}\\

\noindent \textbf{Please explain how it would change your opinion of an organization if you discovered that they were using a TLS proxy.}\\

\noindent \textbf{If you have any other thoughts, please share them with us below:}\\

\section{Second Survey}
\label{appx:study-two}
\vspace{\baselineskip}

\subsection{Page One}

\noindent \textbf{What is your gender?}
\textit{
\begin{itemize}[label={$\circ$},leftmargin=0pt,noitemsep,nolistsep]
\item Male
\item Female
\item I prefer not to answer\\
\end{itemize}
}

\noindent \textbf{What is your age?}
\textit{
\begin{itemize}[label=$\circ$,leftmargin=10pt,noitemsep,nolistsep]
\item 18 -- 24 years old
\item 25 -- 34 years old
\item 35 -- 44 years old
\item 45 -- 54 years old
\item 55 years or older
\item I prefer not to answer\\
\end{itemize}
}

\noindent \textbf{What is the highest degree or level of school you have completed?}
\textit{
\begin{itemize}[label=$\circ$,leftmargin=10pt,noitemsep,nolistsep]
\item Some school, no high school diploma
\item High school graduate, diploma or the equivalent (for example: GED)
\item Some college or university credit, no degree
\item College or university degree
\item Post-secondary education
\item I prefer not to answer\\
\end{itemize}
}

\noindent \textbf{What is your marital status?}
\textit{
\begin{itemize}[label=$\circ$,leftmargin=10pt,noitemsep,nolistsep]
\item Married
\item Single
\item Other 
\item I prefer not to answer\\
\end{itemize}
}

\noindent \textbf{Do you have children?}
\textit{
\begin{itemize}[label=$\circ$,leftmargin=10pt,noitemsep,nolistsep]
\item Yes
\item No
\item I prefer not to answer\\
\end{itemize}
}

\noindent \textbf{In which country do you reside?}
\textit{
\begin{itemize}[label=$\circ$,leftmargin=10pt,noitemsep,nolistsep]
\item United States
\item India
\item Other\\
\end{itemize}
}

\noindent \textbf{How knowledgeable are you about Internet security?}
\textit{
\begin{itemize}[label=$\circ$,leftmargin=10pt,noitemsep,nolistsep]
\item Expert
\item Highly knowledgeable
\item Mildly knowledgeable
\item Somewhat knowledgeable
\item No Knowledge
\item I prefer not to answer
\end{itemize}
}

\subsection{Page 2}

\vspace{\baselineskip}

\noindent \textsc{Ordering of questions randomized.}\\
\noindent \texttt{Strongly Disagree, Disagree, Neither Agree nor Disagree, Agree, Strongly Agree}
\textit{
\begin{itemize}[label=$\circ$,leftmargin=10pt,noitemsep,nolistsep]
\item The above description of TLS proxies helped me to clearly understand what TLS proxies are and how they are used.
\item Stealing passwords and identity theft are in the list of malicious uses shown above.
\item Blocking malware and viruses are in the list of malicious uses shown above.
\item Prior to taking this survey, I was aware that organizations were using TLS proxies.
\item TLS proxies are an invasion of privacy.
\item There are acceptable uses for TLS proxies.\\
\end{itemize}}

\subsection{Page 3}

\noindent For each scenario listed below, provide your opinion on whether or not
the organization should be allowed to run a TLS proxy.\\

\noindent \textsc{Ordering of questions randomized.}\\
\noindent \texttt{No, Only if I consent, Only if I am notified (consent not required), Yes (Neither notification nor consent required), Unsure}
\textit{
\begin{itemize}[label=$\circ$,leftmargin=10pt,noitemsep,nolistsep]
\item Your employer when you use a company computer
\item Your employer when using your own device (cell phone, tablet, laptop)
\item Elementary school
\item Public Library
\item University
\item Paid WiFi -- Airport, Hotel, Cafe, etc.
\item Free WiFi -- Airport, Hotel, Cafe, etc.
\item The company that provides Internet access at your home
\item Personal firewall -- software that you have installed to protect your computer
\item Your government monitoring your Internet traffic
\end{itemize}
}

\subsection{Page 4}

\noindent \textbf{Please feel free to write any thoughts you have on the subject of TLS proxies. We will use this information to help guide future research. (Optional)}

\end{document}